\documentclass{article}

\usepackage{arxiv}
\usepackage{natbib}
\usepackage{subcaption}

\usepackage[utf8]{inputenc} 
\usepackage[T1]{fontenc}    
\usepackage{hyperref}       
\usepackage{url}            
\usepackage{booktabs}       
\usepackage{amsfonts}       
\usepackage{nicefrac}       
\usepackage{microtype}      
\usepackage{graphicx}
\usepackage{fancyhdr}
\usepackage[normalem]{ulem} 
\usepackage{xcolor} 
\graphicspath{ {./images/} }
\newcommand{\mycomment}[1]{}
\usepackage{xcolor}

\title{AAAI-26 Dual Submissions: Novel Challenges}
\author{
 Kiri L. Wagstaff \\
  OSU Libraries\\
  Oregon State University\\
  Corvallis, OR 97330 \\
  \texttt{wkiri@wkiri.com} \\
   \And
 Joydeep Biswas \\
  Computer Science Department\\
  University of Texas at Austin\\
  Austin, TX 78712\\
  \texttt{joydeepb@cs.utexas.edu} \\
  \And
  Erich Merrill III \\
  Independent Researcher\\
  \texttt{erich@erichmerrill.com} \\
  \And
  Bo An\\
  College of Computing and Data Science\\
  Nanyang Technological University\\
  Singapore \\
  \texttt{boan@ntu.edu.sg}
  \And
  Ida Camacho \\
  AAAI \\
  Washington, DC 20004 \\
  \texttt{camacho@aaai.org}
  \And
    David J. Crandall \\
  Department of Computer Science \\
  Indiana University\\
  Bloomington, IN 47405 \\
  \texttt{djcran@iu.edu}
   \And
  Matthew E. Taylor\\
  Department of Computing Science\\
  University of Alberta\\
  Edmonton, AB T6G 2E8 \\
  \texttt{matthew.e.taylor@ualberta.ca}
}

\begin{document}
\maketitle
\begin{abstract}
Dual submissions, in which identical or substantially similar papers are simultaneously submitted to one or more archival venues, without cross-citation or disclosure, are a growing problem for the AAAI Conference and other scientific publication venues.  These submissions increase the burden on the peer-review system and pollute the scientific record.

As part of the AAAI-26 review process, we (conference organizers) compared AAAI main-track submissions to nine other archival venues with overlapping review periods.  We also searched for dual submissions within the AAAI-26 main track.  We employed title+abstract similarity assessment to prioritize highly similar paper pairs for subsequent triage by an LLM-based overlap assessment tool, followed by manual review of the highest severity pairs.  Manual review of such pairs led to the desk-rejection of 141 AAAI-26 main-track submissions.

We seek to alert future organizers, and the broader artificial intelligence research community, to the enormous growth in dual submissions.  The incidence of exact duplicate submissions, which are easy to detect, has been eclipsed by the number of papers that use different words to describe the same contribution, which are extremely time-consuming to detect.   The growth in this phenomenon is likely facilitated by increasing access to generative AI tools.  We include several recommendations for addressing this challenge, including (1) updating the AAAI Multiple Submission Policy and educating the community about acceptable practice, (2) having dual-submission checking tools in place before submissions close, (3) working across venues to converge on consistent policies and penalties to aid in reducing the incidence of dual submission, and (4) creating a community-driven adversarial challenge to accelerate the development of robust detection tools.
\end{abstract}


\section{Introduction}

The AAAI Conference on Artificial Intelligence does not allow dual submissions, in which the same (or similar) manuscript is under review by more than one archival publication venue at the same time.  Unfortunately, there continue to be many violations of this policy, resulting in desk rejections for the manuscripts involved.  Therefore, checking for dual submissions is now a standard part of the compliance checks for submissions to the AAAI Conference (and elsewhere).
While exact duplicate submissions are generally straightforward to detect, recently we have observed a growing trend marked by the submission of highly similar papers that are not verbatim duplicates but that effectively convey the same ideas, contents, and conclusions while using alternative phrasings (see Section~\ref{sec:problem} for details).  Detecting these more subtle kinds of dual submissions has greatly complicated and slowed this part of the conference review process.

For the 2026 AAAI Conference on Artificial Intelligence (AAAI-26), we (the organizers) implemented dual-submission checking by examining pairs of papers that had at least one author in common.  We checked pairs of papers with overlapping authors, both within the AAAI-26 submission pool and for pairs in which one paper was submitted to AAAI-26  and the other paper to another conference or journal whose review period overlapped with that of AAAI-26. Our analysis of AAAI-26 submissions suggests that there is an alarmingly large number (likely thousands) of dual submissions in both the within-AAAI and cross-venue contexts.

The goal of the present report is to elevate this issue to the community's attention due to the recent increase in both the number and the variety of dual submissions.  We urgently need community discussion and development of solutions.  
\paragraph{Why now?} Although dual submissions have occurred in the past, and we were warned by the AAAI-25 organizers to perform checks for them, the widespread availability of generative AI systems that can produce an alternative version of a paper within seconds, with minimal effort, has likely contributed to a staggering growth in this phenomenon.  \citet{bertone:publishless26} warns that these conditions can lead to papers that have ``{\em negative} epistemic value,'' meaning that the knowledge they contribute is outweighed by the time and attention they consume from editors, readers, etc.  This is especially true for dual submissions, which may contribute little or no additional knowledge with respect to each other, while consuming multiples of the same review resources.

In this report, we summarize how we addressed the challenge of dual submissions for AAAI-26, what we found, and the recommendations that we offer to the community.  Dual submissions are likely to increase, both in volume and in variations.  Our community needs to ($i$) discuss and articulate clear dual submission policies (and penalties) and ($ii$) develop capable tools to enforce them. Given the incidence of cross-venue dual submissions, it will be valuable to identify the appropriate level of coordination needed, while preserving the confidentiality of submissions.  A community-wide consensus approach to policies and penalties will provide clarity to authors about what is acceptable publishing behavior and help reduce the number of violations, whether inadvertent or intentional.

\section{Why dual submission is a problem}
\label{sec:problem}

We identified the following categories of dual submissions in the AAAI-26 review process, for submission pairs (A, B) with at least one shared author:
\begin{enumerate}
\item {\bf Exact duplicates:} Submissions A and B are identical.
\item {\bf Obfuscated duplicates:} Submissions A and B present the same solution to the same problem, rewritten to appear different.  Often this manifests as reporting identical empirical results in both papers, attributed to algorithms with different names.

\item {\bf Alternative universe:} Submissions A and B present two slightly different solutions to the same problem.  They evaluate on the same data sets using the same metrics, but the later of the two submissions does not compare to the method in the earlier one (despite joint authorship).  The submissions seem to be derived from the same work and effectively compete with or contradict each other, without citation or acknowledgment.  When both claim to have the ``best'' solution to the problem, at most one of them can be true.

\item {\bf ``Salami'' slices:} Submissions A and B present highly related elements of the same system or solution, often with significant portions of overlapping text.
\end{enumerate}

There are several reasons why these dual submissions are a problem for the research publication process:
\begin{enumerate}
\item Violating a dual submission policy breaks our expectations of {\bf scientific integrity}.  If a publication venue disallows dual submission (as most peer-reviewed and archival venues do), then submitting to two venues violates that rule.  Publishing a technical advance in two venues would make proper citation of the work ambiguous and problematic, especially for obfuscated duplicates.  For alternative universe submissions, it is intellectually dishonest to submit one solution and omit a comparison with a known superior solution. This applies whether the superior solution was contributed by the same authors or by other authors, but it is especially egregious for dual submissions, where the authors cannot claim ignorance of the superior method.
\item Alternative universe submissions represent {\bf an unreliable approach to science} that relies on the peer review system itself to decide between different solutions to the same problem.  Instead, authors ought to use their own judgment and evaluations to make that initial comparison between known solutions, then only submit work that presents their best result (or carefully disclose and compare the two ideas and their relative merits).
\item Dual submissions, even when they are not exact duplicates, can {\bf pollute the scientific record.} Obfuscated duplicates and alternative universe submissions enable a shotgun-like approach to publication in which a single source produces multiple possible manuscripts to achieve the goal of publishing at least one of them.  Given the inherent stochasticity in the peer review system, and the fact that the submissions are reviewed by different reviewers, this is not a reliable, stable, or objective way to advance the field or find the best solution to a given problem. 

\item Dual submissions challenge our current model of peer review which assesses {\bf each manuscript independently}, relying on the authors to provide key context.  When we cannot trust authors to disclose known alternative solutions, we must invest additional effort in cross-comparing multiple submissions to assess whether they are distinct contributions.
\item Dual submissions {\bf overburden the already strained peer review system}.  AAAI-26 received 28,000 submissions, versus the 12,000 submissions to AAAI-25, demanding the recruitment of twice as many reviewers as expected.  Review quality suffers.  Additional organizer effort is required to check for and address violations of the dual submission policy.
\end{enumerate}

\section{Handling dual submissions at AAAI-26}
For AAAI-26, we followed the AAAI-26 Multiple Submission policy\footnote{\url{https://aaai.org/conference/aaai/aaai-26/submission-instructions/}} provided to authors, which disallowed the submission to AAAI of manuscripts already under review or published by an archival venue, as well as the subsequent submission of a AAAI submission under review to another such venue:

\begin{quote}
``AAAI-26 will not consider any paper that, at the time of submission, is under review or has already been published or accepted for publication in any archival venue such as a journal or a conference (workshops and preprint servers such as arXiv are acceptable). Authors are free to retract a submission from a venue with a concurrent review process (e.g., from NeurIPS-25) and submit the same work to AAAI-26 as a regular paper, provided that this retraction occurs before the AAAI-26 submission. Authors must confirm at the time of submission that the paper is not under review at another archival conference or journal.

Once they have made a submission to AAAI-26, authors may not submit the same paper to another archival conference or journal until they receive an accept/reject decision from AAAI-26 or they withdraw their submission from AAAI-26. In some cases, it may require a judgment call to determine whether two concurrent submissions constitute a violation of AAAI's multiple submission policy. If a concern is raised about the similarity of two non-identical submissions, at least three people will inspect whatever information is available about both submissions. If they all agree that the simultaneous submission has excessive technical overlap, the paper will be summarily rejected and the organizers of the other conference will be informed about AAAI's decision. As with all summary rejects, such decisions are final.''
\end{quote}

We focused on pairs of submissions with at least one author in common.  We checked for dual submissions by comparing AAAI-26 submissions to those received by other conferences whose review period overlapped with that of AAAI-26 (``cross-venue'' pairs).  These venues included AAMAS, ARR (ACL Rolling Review), EMNLP, ICLR, KDD, NeurIPS, 
and WACV, as well as AIA and IAAI (co-located with AAAI-26). We also checked for dual submissions within the set of manuscripts submitted to AAAI-26 (``within-AAAI'' pairs).

\begin{figure}
  \centering
  \includegraphics[width=\linewidth]{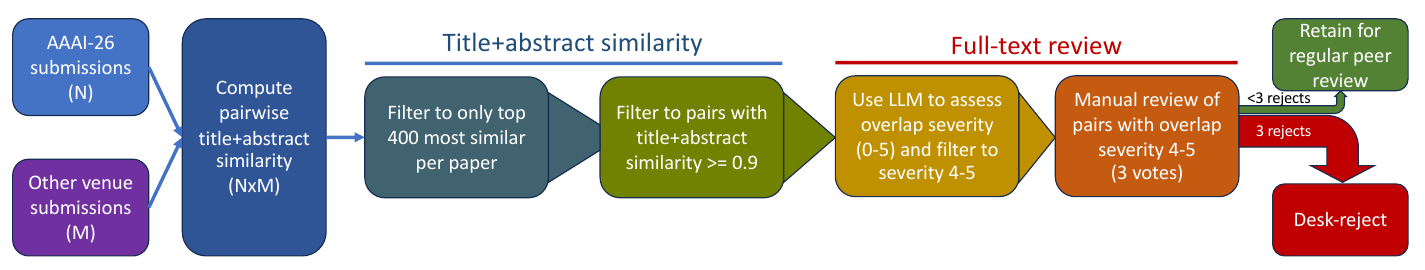}
  \caption{The dual submission review process employed for AAAI-26.  Two collections of papers are provided as input (the second venue could be AAAI-26 itself for within-conference comparison).  Pairwise similarity scores are computed based on title+abstract information, then an LLM is used to assess full-text overlap, and finally candidates are reviewed manually by three humans.  Each stage greatly reduces the number of candidates that move to the next one.}
  \label{fig:system}
\end{figure}

Our approach to detecting and checking for dual submission at AAAI-26 consisted of three steps, illustrated in Figure~\ref{fig:system}: ($i$) identify pairs of submissions with overlapping abstract content; ($ii$) assess overlap severity with an LLM-based comparison of the full-text content; and ($iii$) manually review pairs of manuscripts with the highest overlap severity to determine whether they violated the AAAI-26 submission rules.
At every step, we sought to keep information sharing between venues to the minimum necessary so as to protect the confidentiality of authors and their contributions as much as possible.

\subsection{Identify dual submission candidates}
We began by comparing paper meta-data (author, title, and abstract) to identify candidate pairs of concern.  To identify cross-conference pairs, we established agreements with other venues to allow OpenReview staff to compute pairwise similarity based on the title and abstract, for paper pairs with at least one overlapping author.  Entrusting this step to OpenReview meant we did not need access to the complete set of submissions to other venues, nor to share such information about AAAI submissions with them.

OpenReview staff computed pairwise title+abstract similarity, using the SPECTER2+SciNCL model, for all pairs of papers.  SPECTER2~\citep{singh-etal-2023-scirepeval} uses a SciBERT-based document encoding learned from scientific texts in which training examples were chosen using  citation relationships as a strong similarity signal. SciNCL~\citep{ostendorff-etal-2022-neighborhood} instead learns an embedding of citation relationships, then chooses the $k$ nearest neighbors in this space to train the document embedding model.  Each model computes pairwise similarity as the dot product between the vector embeddings of the two papers, then normalizes all scores in a batch to fall between 0 and 1.  SPECTER2+SciNCL uses the average of these two similarity scores, which are first individually normalized to the range $[0, 1]$. 
%
OpenReview then truncated the pairwise score list to include only the top 400 matches (by similarity score) for each paper, then filtered the list to only those pairs with at least one overlapping author.  

\subsection{Check full-text overlap for high-similarity pairs}
While high abstract similarity presented an initial concern, we needed to look at the full text of each pair of manuscripts to make a final determination about problematic similarity.  Some manuscripts might have similar abstracts yet provide genuinely independent contributions when the full text was considered.  We forged individual cross-conference agreements to allow the comparison of a select group of anonymized PDFs, decoupled from their submission meta-data, for only those manuscripts with extremely high abstract similarities.

Given the enormous number of submission pairs with high abstract similarity, it was obvious that we could not manually check all of the full-text PDFs (or even their abstracts).  We therefore turned to a large language model (GPT-5 from OpenAI) to assist in triaging the pairs by analyzing the full-text, anonymized PDFs and directing our attention to the submissions with most severe overlap.  This LLM used a zero data retention policy, so no submission information would be visible to or retained by OpenAI.

For manuscript pairs with a title--abstract similarity of at least $0.9$ (an arbitrary but useful threshold), we prompted the LLM to compare their anonymized PDFs and assign each pair an ``overlap severity'' score, which ranges from 0 to 5, together with a short prose explanation justifying the score. The complete prompt, including definitions of the overlap severity score values, is shown in Appendix~\ref{app:LLM}.  We did not provide the LLM with any further information about the pair of manuscripts (neither their abstract similarity score, nor other metadata).

Operationally, we pre-uploaded the anonymized PDFs to OpenAI and used a file list mapping each \texttt{openreview-id.pdf} filename to an OpenAI file identifier.  The batch comparison tool then processed candidate pairs in parallel and wrote incremental CSV output containing the two submission IDs, their OpenReview forum links, the upstream title-abstract similarity score, the LLM overlap severity score, and the LLM explanation.

Within the set of manuscripts submitted to AAAI, pairs that received an overlap severity score of 4 (``extensive overlap'') or 5 (``clear plagiarism'') were chosen for further scrutiny.  We sent a warning message to the authors of each such AAAI submission notifying them that their manuscript may be subject to closer review due to its potential overlap with another manuscript, that they were encouraged to withdraw any too-similar submissions, and that in the future, dual submission violations would lead to additional penalties.

We did not use the LLM to make any decisions about whether a pair of manuscripts violated the Multiple Submission Policy.  This tool was experimental and has not yet been validated against human judgments (but it should be, as we will need a tool like it in the future).  Instead, we used its output only to prioritize manuscripts for further manual review.

\subsection{Perform manual review of top-severity pairs}
AAAI-26 policy required that three people inspect each non-identical submission pair of concern before reaching a decision to reject such a pair. In general, the submissions had not been assigned the same set of reviewers, or even the same Area Chair, so manual review had to be done independently by the Program Chairs, Associate Program Chairs, Ethics Chairs, and some additional recruits.  To follow our stated policy, three humans checked each manuscript pair that was assigned overlap severity 5 (and some severity 4 as time permitted).  We provided the same training and examples to each dual reviewer to support consistency in assessment.  This process was extremely time-consuming, during a period that is already very high workload for conference organizers.  To limit information sharing, we carefully restricted visibility of the anonymized PDFs to only three organizers.  If all three humans concurred that the manuscript violated the Multiple Submission policy, the submission was desk-rejected and the authors were notified of the reason for this decision.  For cross-venue pairs, we also notified the other venue and offered to share the corresponding anonymized AAAI submission so they could reach their own accept/reject decision for the manuscript in their submission pool.  

\section{Results}

After removing empty, overlength, incorrectly formatted, and non-anonymized submissions, AAAI-26 had 22,423 manuscripts under review. 

The first venue we discussed dual submissions with was NeurIPS-25, whose decision deadline (September 18) came before that of AAAI-26 (November 8).  NeurIPS staff had computed their own edit distances between NeurIPS and AAAI submissions based on title and abstract and found 47 exact duplicates and 9 highly similar manuscript pairs.  We reviewed the titles and abstracts of the 47 duplicates and concurred.  Checking the PDFs revealed that one of the duplicate abstracts had an invalid AAAI-26 PDF so was likely a test submission and would not be penalized, yielding 46 actual duplicates that were desk-rejected.  Three AAAI-26 organizers independently examined the full text PDFs for the 9 highly similar pairs; unanimous votes resulted in desk-rejecting 8 of them.  The remaining submission was deemed a ``salami'' slice but allowed to proceed in review.
We also applied our own process for assessing title+abstract similarity (described above) to AAAI-NeurIPS pairs. NeurIPS granted us access to the anonymized PDFs for the top 100 AAAI-NeurIPS pairs (based on SPECTER2+SciNCL scores).  This led to the rejection of an additional 9 AAAI (and NeurIPS) submissions.

We contacted eight additional venues with overlapping review periods and worked with OpenReview to generate pairwise title+abstract SPECTER2+SciNCL similarity score lists for each one. Filtering to the top 400 pairs per paper resulted in:
\begin{itemize}
    \item 98,846 cross-venue pairs with at least one shared author
    \item 17,170 within-AAAI pairs (main track) with at least one shared author (13,406 unique AAAI papers)
\end{itemize}
Filtering each list to only those pairs with at least $0.9$ similarity resulted in:
\begin{itemize}
\item 11,475 cross-venue pairs with at least one shared author
\item 6,061 within-AAAI pairs (main track)
  \end{itemize} 

\begin{figure}
  \centering
  \includegraphics[width=0.5\linewidth]{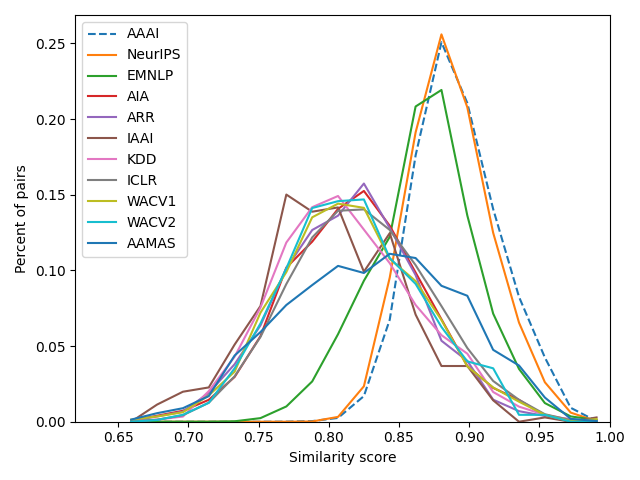}
  \caption{Distribution of pairwise abstract similarity scores between manuscripts submitted to AAAI-26 and other venues, for the top 400 matches per submission, then filtered to those pairs with at least one shared author.  Since scores are separately normalized for each venue, they are not directly comparable across venues.}
  \label{fig:scores}
\end{figure}

The distribution of pairwise abstract similarity scores, by venue, is shown in Figure~\ref{fig:scores}.  We found the highest similarity for co-authored submissions that were submitted simultaneously to AAAI-26's main track.
Since scores are separately normalized for each venue, they are not directly comparable across venues.  However, we found it useful to assess the number of pairs that scored at least $0.9$ similarity for each venue (i.e.,~the top 10\% of similarity score range for that venue). Table~\ref{tab:res} tallies the number of co-authored pairs, how many (and what percentage) had a similarity of at least $0.9$, how many received manual review, and how many resulted in the rejection of a AAAI-26 main track paper.  Venues with more than 10\% of their co-authored pairs scoring at least $0.9$ similarity are highlighted in \textcolor{red}{red}.  These venues include AAMAS, EMNLP, and NeurIPS.  Within AAAI-26, 35\% of co-authored pairs had at least $0.9$ similarity.

{\bf A total of 141 AAAI main track submissions were desk-rejected for violating the AAAI-26 Multiple Submission policy.}  Based on our findings, the true number of violating submissions is likely much, much higher.  We did not have the time or effort available for three organizers to manually review every pair of concern.

\begin{table}
  \caption{Pairs of manuscripts analyzed (those with at least one overlapping author).  Venues with more than 10\% of co-authored pairs scoring at least $0.9$ title+abstract similarity are highlighted in \textcolor{red}{red}. Submission pairs with exact duplicate abstracts did not require manual ({\bf full-text}) review, so the number of submissions rejected can exceed the number receiving manual review. The table reports the venue-level shared-author accounting; the final desk-rejection total includes duplicate submissions found without shared authors and appeal-stage adjustments, as discussed below.}
    \label{tab:res}
  \centering
  \begin{tabular}{|l|rrrr|r|} \hline
  Venue & Co-authored pairs & \multicolumn{2}{c}{$\geq 0.9$ sim (\%)} &  Manual review & AAAI papers rejected \\ \hline
  AAAI--AAMAS & 2,126 &   289 & \textcolor{red}{(14\%)} & 3 & 2 \\
  AAAI--AIA   & 1,519 &    78 & (5\%)  & 4 & 2 \\
  AAAI--ARR   & 3,157 &   139 & (4\%)  & 4 & 6 \\
  AAAI--EMNLP & 7,133 & 1,228 & \textcolor{red}{(17\%)} & 5 & 11 \\
  AAAI--IAAI  &   353 &    11 & (3\%)  & 5 & 1 \\
  AAAI--ICLR  & 54,542 & 3,612 & (7\%) & 26 & 18 \\  
  AAAI--KDD   & 3,509  &   185 & (5\%) & 4 & 4 \\
  AAAI--NeurIPS & 18,759 & 5,721 & \textcolor{red}{(30\%)} & 9 & 57* \\ 
  AAAI--WACV Round 1 & 2,870 &  160 & (6\%) & 4 & 4 \\
  AAAI--WACV Round 2 & 878  &   52 & (6\%) & 0 & 0 \\ \hline
  Total cross-venue & 98,846 & 11,475 & (12\%) & 64 & 105 \\
  \hline \hline
  Within AAAI main track & 17,170 & 6,061 & \textcolor{red}{(35\%)} & 27 & 36 \\ \hline
\end{tabular} \\
 {\small *The 57 NeurIPS rejections include 3 exact duplicates from the NeurIPS Datasets \& Benchmarks track.}
\end{table}

\subsection{Analysis of title+abstract similarities within the AAAI main track}

Initially we were only looking for cross-venue dual submissions.  However, after noticing that there were two cases in the nine ``highly similar'' pairs of concern identified by NeurIPS in which a single NeurIPS paper had high similarity with {\em two} AAAI-26 main track submissions, we realized that we also needed to look for possible dual submissions within the AAAI-26 main track submission pool itself.  
This analysis proved to yield additional insights.

\begin{figure}
    \centering
    \begin{subfigure}[b]{0.45\textwidth}
        \centering
        \includegraphics[width=\textwidth]{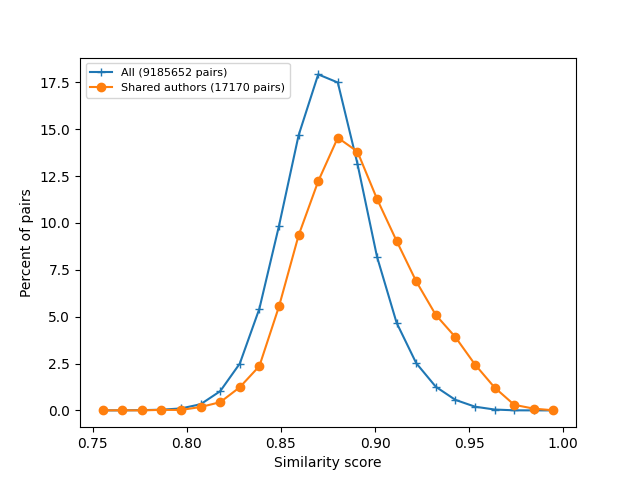}
        \caption{Pairwise similarity of all manuscripts (close to a Gaussian distribution) versus those with shared authors (shows non-Gaussian features).}
    \end{subfigure}
    \hfill
    \begin{subfigure}[b]{0.45\textwidth}
        \centering
        \includegraphics[width=\textwidth]{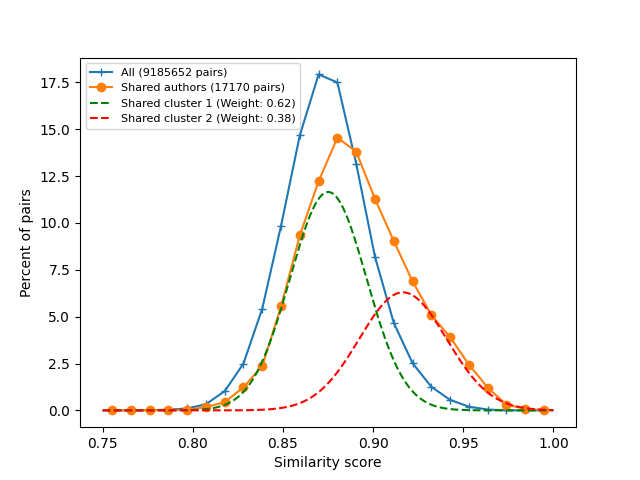}
        \caption{A two-component Gaussian mixture model (dashed curves) helps explain the non-Gaussian shape of the shared-authors distribution.}
    \end{subfigure}
    \caption{Distribution of pairwise title+abstract similarity scores for manuscripts simultaneously submitted to AAAI-26 ($n=22,423$ submissions). Initially, scores were computed for all possible pairs, but these were filtered by OpenReview to only the top 400 most-similar pairs per manuscript, which are the scores included in the ``All'' curve.}
    \label{fig:aaai-sim}
\end{figure}

While title+abstract similarity scores cannot be directly compared between venues (as noted above), it is valid to compare scores for different subsets of a given run, such as the all-pairs analysis of AAAI-26 main track submissions.  We found that submission pairs with at least one shared author tended to have higher similarity scores (see Figure~\ref{fig:aaai-sim}(a)).  (Note: while these scores were initially computed for all 502 million possible pairs, OpenReview retained only the top 400 most-similar pairs for each manuscript, so the ``All'' curve contains only 9 million pairs, and both curves are shifted higher than the distribution of similarity would be for randomly chosen pairs of manuscripts.) Elevated similarity for shared-author pairs need not signal a problem on its own; manuscripts that have shared co-authors are more likely to be written by researchers that have similar interests, and the shared co-author(s) may have contributed to the writing of both submissions, increasing the similarity of the manuscripts' writing style.

However, the ``Shared authors'' curve also differs from the ``All'' curve in that it exhibits non-Gaussian features, including an elevated proportion of the highest-similarity pairs.  We hypothesize that there were multiple sub-populations within this group.  We performed Gaussian mixture modeling on just the shared-author paper scores, with two components, and obtained the dashed curves shown in Figure~\ref{fig:aaai-sim}(b).  Cluster 1, representing 62\% of these pairs, has a mean of 0.875, very close to the mean ``All'' score of 0.874.  Cluster 2 represents 38\% of the shared-author pairs, with a distinctly elevated mean similarity score of 0.917, which is very unlikely to occur in the overall data set; only 4.6\% of all pairs are that similar.  This anomalous component calls for additional scrutiny of such pairs.  Interestingly, the Bayes optimal threshold for the decision boundary between these components is 0.903, almost exactly the criterion we (heuristically) used to decide which pairs would be further assessed by the LLM overlap severity tool between multiple venues.

We applied the LLM overlap severity assessment to {\em all} co-authored pairs of AAAI-26 main track papers, regardless of similarity score.
As described above, we sent warning messages to authors of submissions judged as overlap severity 4 or 5, to deter authors and reduce our reviewing burden.  Ideally, this assessment should be performed immediately after submission.  Because it took some time for us to discover the extent of the problem and to develop tools to investigate it, the warning messages were not sent until right around the time that AAAI Phase 1 rejection decisions (based on regular peer review) went out.  Therefore, we do not have a good way to determine the efficacy of this warning.  We had identified 298 pairs (545 unique manuscripts) of AAAI-26 submissions of concern (overlap severity 4 or 5).  As of September 15, 2025, 318 of these 545 submissions were already withdrawn or rejected by Phase 1, so warning messages were sent only to the authors of the 227 remaining submissions. However, significant reviewer effort had already been expended on the other 318 manuscripts by that point.

\subsection{How did the LLM overlap severity tool perform?}

\begin{table}
  \caption{LLM overlap severity score results.  In most cases, all pairs with title+abstract similarity of at least $0.9$ were analyzed (except those that had already been rejected or withdrawn by that point).  Exceptions: Only the top 100 pairs (instead of 5,721) for NeurIPS and top 21 pairs (instead of 1,228) for EMNLP.
  WACV declined to participate in this analysis step.  We analyzed all pairs of AAAI papers with shared authors, plus a similar number of AAAI pairs with no shared authors that had the highest scores, for comparison.  ``Severity 5 precision'' is calculated as the fraction of pairs scored by the LLM as overlap severity 5, and manually reviewed, that were rejected after human review. Only 21 of 41 severity-5 AAAI-AAAI pairs were manually reviewed because by that point the rest had already been rejected by peer review. Counts in this table exclude failed LLM comparisons. For the AAAI main-track shared-author row, 17,205 comparisons were attempted and 435 failed, primarily due to missing PDFs in the LLM's file storage.}
    \label{tab:llmres}
  \centering
    \begin{tabular}{|l|r|rrr|r|} \hline
  & Pairs (full-text) & \multicolumn{3}{c|}{LLM overlap severity} & Severity 5 precision \\
  Venue & analyzed by LLM & 0-3 & 4 & 5 & (from manual review) \\ \hline
  AAAI--AAMAS & 288 & 273 & 12 & 3 & 2/3 = 0.67 \\
  AAAI--AIA   &  78 &  75 &  3 & 0 & N/A \\
  AAAI--ARR   & 139 & 131 &  4 & 4 & 4/4 = 1.00 \\
  AAAI--EMNLP &  21 &  14 &  3 & 4 & 4/4 = 1.00 \\
  AAAI--IAAI  &   5 &   2 &  2 & 1 & 1/1 = 1.00 \\
  AAAI--ICLR  & 3,599 & 3,445 & 128 & 26 & 18/26 = 0.69 \\ 
  AAAI--KDD   & 182 & 176 &  4 & 2 & 2/2 = 1.00 \\
  AAAI--NeurIPS & 100 & 55 & 36 & 9 & 8/9 = 0.89 \\
  AAAI--WACV  &  0 & N/A & N/A & N/A & N/A \\
  AAAI--WACV2 &  0 & N/A & N/A & N/A & N/A \\ \hline
  Total cross-venue & 4,412 & 4,171 & 192 & 49 & 39/49 = 0.80\\ 
  \hline \hline
  AAAI main track (shared authors) & 16,770 & 16,472 & 257 & 41 & 12/21 = 0.57 \\
  AAAI main track (no shared authors) & 18,250 & 18,209 & 37 & 4 & 3/3 = 1.00 \\ \hline
\end{tabular} \\
\end{table}

The LLM overlap severity tool was extremely useful for deciding which submissions to prioritize for the limited manual review time that we had available.  In addition to drawing our attention to the most problematic pairs, the LLM also provided a rationale for each pair that flagged specific sections with overlapping text, the same methodology and evaluation, and/or results that were identical to the last decimal place shown.

Although we did not instruct the LLM about what distribution of scores to assign, we typically found, across all venues, that 1.1\% of pairs with title+abstract similarity $\geq 0.9$ were deemed severity 5, and an additional 4.4\% were deemed severity 4. See Table~\ref{tab:llmres} for full numbers.  The within-AAAI shared-author run had slightly lower rates among those with title+abstract similarity $\geq 0.9$, with 41 severity-5 pairs (0.8\%) and 257 severity-4 pairs (4.2\%), but otherwise tracked the same trend.

We also found that the LLM overlap severity score had high precision, i.e.,~papers that were assigned an overlap severity 5 were generally rejected after manual review.  However, we suspect that recall was rather low; separate review of some highly similar pairs led to rejection despite the LLM assigning an overlap severity of only 2.
This supported our choice to treat the LLM overlap severity assessment as a prioritization tool rather than an adjudication tool.

\begin{figure}[t]
  \centering
  \includegraphics[width=4in]{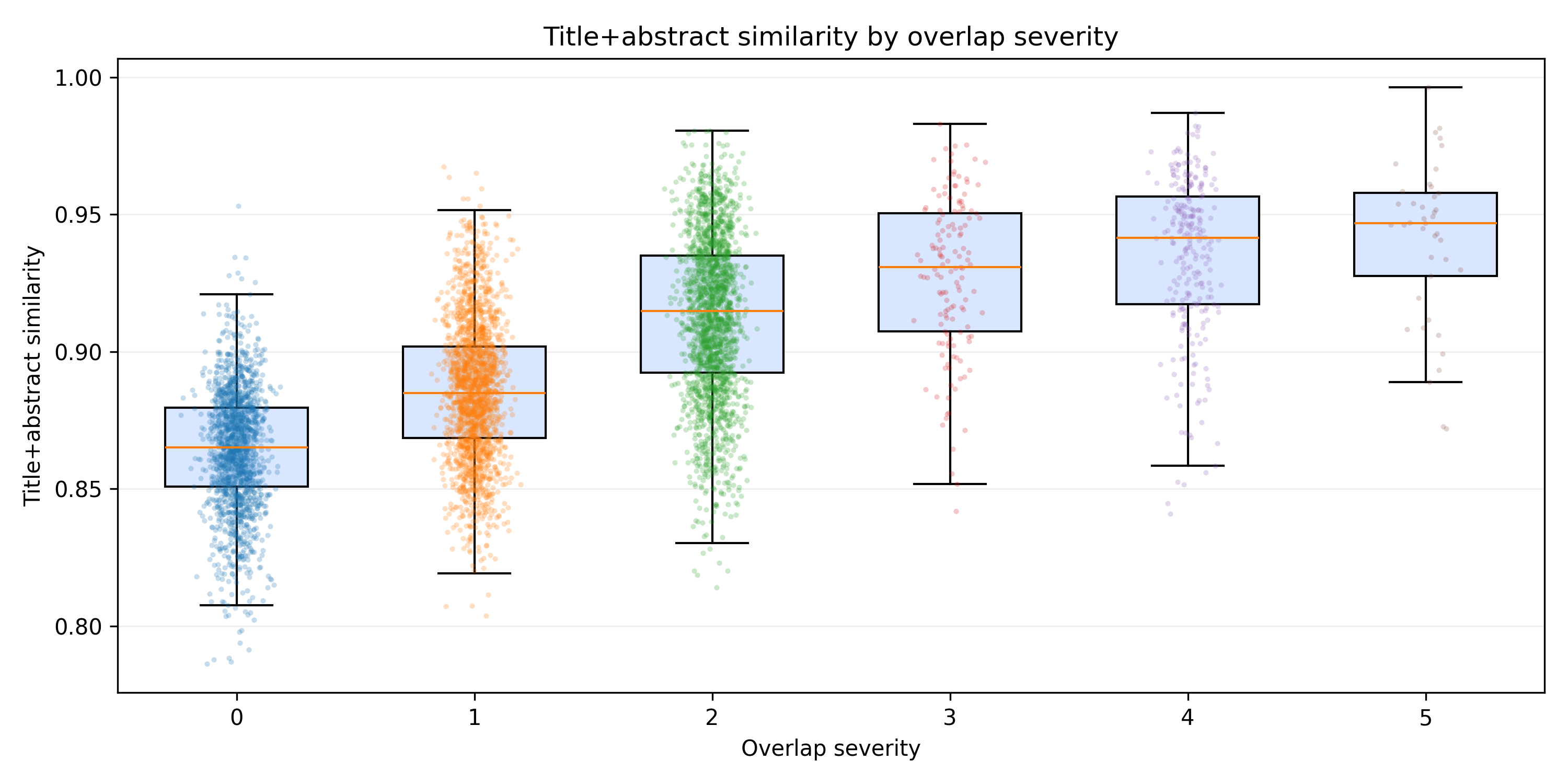}
  \caption{Relationship between the OpenReview title/abstract similarity score and the LLM full-text overlap severity score for within-AAAI shared-author pairs. Failed LLM comparisons are excluded.}
  \label{fig:llm-similarity}
\end{figure}

We were interested in how much the abstract similarity score correlated with the full-text LLM overlap severity assessment.  We found only moderate correlation: excluding failed LLM comparisons, the post-processing analysis found Spearman's $\rho=0.494$ and Pearson's $r=0.508$ between the OpenReview similarity score and the LLM overlap severity score.  The within-AAAI shared-author run had 435 failed comparisons out of 17,205 attempts (2.5\%), almost all due to missing PDF file identifiers.  However, the two assessments are different in two major ways: abstract vs.~full-text and (deterministic) embedding vector similarity vs.~(probabilistic) LLM assessment.  Figure~\ref{fig:llm-similarity} shows substantial spread within each severity level, reinforcing the need for full-text comparison before prioritizing organizer review.

\begin{figure}
  \centering
  \includegraphics[width=5in]{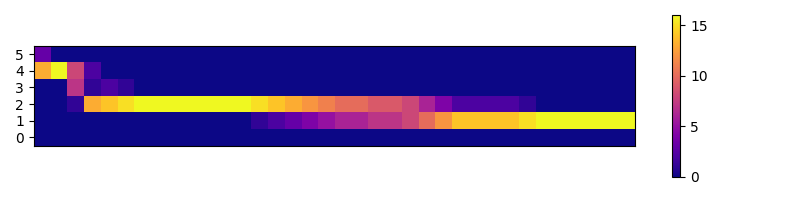}
  \includegraphics[width=5in]{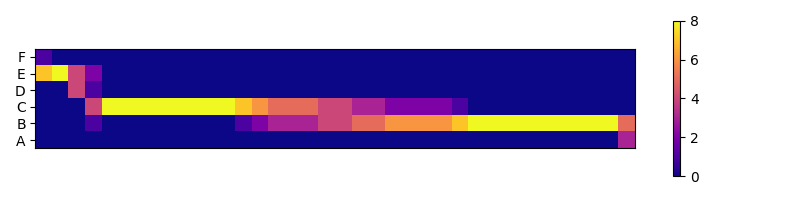}
  \caption{Stability of LLM overlap severity scores (37 pairs, one column per pair, sorted by decreasing severity).  (Top) Heatmap of the distribution of LLM severity scores (0-5) over 16 runs.
  (Bottom) Heatmap with prompt modified to assign symbolic scores (A-F) (8 runs).}
  \label{fig:llm-runs}
\end{figure}

During initial testing and application of the LLM overlap tool, we observed that it occasionally assigned different scores to the same submission pair when executed multiple times. To evaluate the stability of the tool, we selected 37 submission pairs above the similarity threshold and performed 16 independent runs of the LLM overlap tool on these pairs.  Figure~\ref{fig:llm-runs} (top) shows the resulting distribution of scores for each pair as a column-wise heatmap, with one column for each pair.  Most pairs maintained the same score (yellow), while some varied between adjacent scores (reds).

Additionally, we considered that the prompt being used to direct the LLM's scoring and justification may be introducing unintended bias associated with the numeric range used (0--5).  We ran a similar test with a modified prompt that replaced the numeric overlap severity categories (0--5) with symbolic labels (A--F). Over 8 runs on the same paper pairs, the LLM's behavior was roughly the same as when using the numeric labels (see Figure~\ref{fig:llm-runs} (bottom)).

We determined that the LLM tool was consistent enough for our immediate needs, and we proceeded  with using a single execution of the tool to judge each submission pair for the rest of the detection process.
Note, these results do not fully validate the LLM's reliability in this setting, but we lacked the resources (and especially time) to perform a more comprehensive evaluation.  Further evaluation, including a comparison with human overlap severity scoring of the same pairs, is needed.

\section{Lessons and Recommendations}

While there have been reports of and admonishments about problematic dual submissions for decades across many fields of scientific research~\citep[e.g.,][]{bowyer1999multiple,stone-duplicate03,leopold-duplicate13,emanuele-duplicate25}, we found that it was a much larger problem than we anticipated in this conference cycle.  It is likely to continue to grow unless we take rapid steps to make clear that these strategies will not be rewarded.

\paragraph{1. Make Dual Submission Policies Clear and Easy to Follow.}
AAAI-27 has updated its Multiple Submission Policy to {\bf (1A) explicitly disallow obfuscated duplicates and alternative universe submissions}.  The Call for Papers\footnote{https://aaai.org/conference/aaai/aaai-27/main-technical-track-call/} states that 
``Submissions that are deemed substantially similar, for example, multiple papers addressing the same problem with only minor variations in approach, may be desk-rejected at the discretion of the Area Chairs.''  However, a clear statement is needed to {\bf (1B) explicitly disallow
simultaneous submission to other archival venues.}  We are encouraged by recent updates to similar policies by other high-volume AI and machine learning conferences, including NeurIPS and ICML:

\begin{quote}
    {\em NeurIPS 2026 Main Track Handbook}\footnote{\url{https://neurips.cc/Conferences/2026/MainTrackHandbook}}:
``The reviewing process will treat any other archival submission by an overlapping set of authors as prior work (dual submissions to nonarchival workshops are permitted). If publishing one would render the other too incremental, both may be rejected. This includes ``thin slicing'' by submitting two or more very similar papers to NeurIPS in hopes one will be accepted, as well as dual submissions of the same paper to both the main and E\&D track.''
\end{quote}

\begin{quote}
{\em ICML 2026 Call for Papers}\footnote{\url{https://icml.cc/Conferences/2026/CallForPapers}}:
``Authors may not submit papers that are identical, or substantially similar, to versions that have been previously published, accepted for publication, or submitted in parallel to other conferences or journals. Such submissions violate our dual submission policy, and the organizers have the right to reject such submissions, or to remove them from the proceedings. Any concurrent ICML submissions with an overlapping set of authors will also be treated as prior work (so, for example, if publishing one would render the other too incremental, then this may be considered grounds for rejection).''
\end{quote}

We found that several authors were confused or surprised that their dual submissions violated our policy.  We recommend that AAAI (and the larger AI/ML community) {\bf (1C) provide more education about dual submission policies and scientific integrity}.

To further increase awareness and compliance, we recommend that AAAI {\bf (1D) include a statement that authors must agree to at the time of submission, affirming that (A) the manuscript (or substantially similar versions of it) is not already under review in another archival venue, and (B) any other relevant simultaneous submissions to the AAAI main track by any authors of the manuscript are cited in this manuscript, in anonymous fashion as ``under review'' work.}  This statement can include a reminder that violations of this policy can lead to desk-rejection.

\paragraph{2. Process Improvements.}  We identified several recommendations to improve the dual submission assessment process.

{\bf 2A. Designate one of the Associate Program Chairs to oversee the dual submission review process}.  This is a nontrivial job that requires careful attention and management of checking for dual submissions, organizing manual review efforts, and communicating carefully with the authors involved.

{\bf 2B. Have dual-submission checking tools and policy in place before the conference submission deadline.}  For AAAI-27, the abstract deadline is July 21 and full papers are due on July 28.  Checking for dual submissions within the AAAI pool can be automatically run as soon as submissions close, and checking against the papers in other venues can be done in the following week.  This will give the organizers  time to quickly reject clear violations, thereby reducing reviewer burden and sending a clear signal that dual submissions are not allowed and will be caught.  But this fast action will only be possible with reliable tools in place for immediate use.  

Ideally, there should be ongoing work on tool improvements, such as:
\begin{enumerate}
    \item We recommend {\bf choosing or developing an abstract similarity score that can be compared across venues} (and for runs with different subsets of the submission pool).  This could be achieved by turning off SPECTER2+SciNCL score normalization, or using cosine similarity instead of the dot product, since it has a well defined range and would not require further normalization.  
    \item We recommend further {\bf validation and testing (and refinement) of the full-text LLM overlap scoring tool} to determine how best it can be used for triage/prioritization.  The moderate correlation between title/abstract similarity and full-text severity, the imperfect precision of severity-5 cases, and the nondeterminism of LLM outputs all point to the need for calibrated thresholds, repeated-assessment studies, and human review of severe cases.
\end{enumerate}

{\bf 2C. Reduce the number of human organizers (currently three) that must review every dual submission pair of concern that are not exact duplicates}.  With sufficient training for consistency in judgment, plus reliable tools to quickly highlight problematic areas of overlap for human examination, a single organizer could review these candidates and make decisions.  The Multiple Submission Policy should be updated to reflect the requirement that is decided upon.

{\bf 2D. Improve communication with authors.} 
In the ``excessive technical overlap'' warning message, we did not promise to overlook dual submissions if authors withdrew one of their overlapping papers, but many authors made this assumption.  In cases where they withdrew one paper, but we later confirmed there had been a dual submission, we rejected the remaining paper as well, since the dual submission violation had been confirmed and the harm to the system (additional burden) had already occurred.  Some authors protested this decision on the grounds that they had withdrawn one of their dual submissions.  The AAAI Multiple Submission policy, and any subsequent communication to authors, should make the policy (and possible actions) more clear.

{\bf 2E. Aim to minimize the number of papers that are rejected after acceptance, but be willing to take this action when necessary (and inform authors in the AAAI Multiple Submission Policy that this can happen).}  For AAAI-26, a handful of papers were found to be dual submissions after they had already been accepted to the conference.  For consistency with our policy, we had to reject them, which was a painful act for organizers and authors both.  Ideally, this would not happen, and following the first suggestion above will help make it less likely, but if evidence comes to light even after decisions have gone out, it should be assessed with the same standard, akin to retractions made by journals when papers are found to have severe problems post-publication.

\paragraph{3. Community Coordination.}

We recommend that AAAI {\bf (3A) develop standing agreements with other conferences} for secure, minimal information sharing that enables dual submission checking while preserving author/paper confidentiality.  This will reduce the work required to facilitate the cross-venue dual submission checking with each conference cycle.

Ideally, the artificial intelligence research community will {\bf (3B) develop consensus policies and penalties} so that authors have clear guidance and expectations for their submissions.
It will be valuable to discuss a range of penalties so that conferences have options beyond simply rejecting individual papers that violate the policy.  Not every case is the same, and some might require lighter or heavier penalties.  For example, arXiv has the following options available to address problematic author behavior: a submission rate limit, a temporary ban on submissions, or a lifetime submission ban.
Beyond simply rejecting a submission, conferences may consider options such as future submission bans (of varying lengths) to their own venue and whether the penalty applies to only the submitting author or all co-authors.  If a pattern of violations is observed to originate from an institution, actions such as notifying the institution and encouraging remedial author training, or even an institutional ban, may be appropriate.  The most egregious cases might call for coordination between venues so the problem is not simply pushed from one place to another.

There is no single solution to this problem, and we invite community feedback and discussion of this important topic.

\paragraph{4. Launch a Community-Driven Adversarial Challenge.}

We recommend establishing an open, academic competition—akin to a cybersecurity Red Team vs. Blue Team framework—to proactively accelerate the development of robust detection tools.

The Generation Track (Red Team): Researchers would be tasked with deploying advanced AI approaches, such as orchestrated multi-agent systems, to automatically generate sophisticated dual submissions (e.g., ``obfuscated duplicates'' or ``alternative universe'' variations). To preserve the confidentiality of the peer review process, this track should be grounded entirely in a dataset of already-published, open-access papers. This allows the community to safely simulate how bad actors might misuse AI without exposing real, under-review manuscripts.

The Detection Track (Blue Team): Participants would focus on developing novel algorithms to identify these synthetic dual submissions. The objective is to push the community beyond superficial title+abstract text matching to build scalable systems capable of recognizing underlying structural, methodological, and logical overlaps across massive submission pools.

Structuring this challenge as an international collaboration across top universities and conferences would maximize its value and execution. By pooling global institutional resources, the AI community can rapidly crowdsource solutions to a problem that inherently threatens the broader scientific record.

\section*{Acknowledgments}

We thank everyone who helped read, review, and check for dual submissions.  In addition to those included as authors of this report, this dauntless crew includes David Crandall, Odest Chadwicke Jenkins, Matthew Lease, and Dylan Shell.  We also thank the organizers of other venues with overlapping review periods who were excellent partners in the process, the AAAI-25 organizers who first alerted us to this problem, and OpenReview staff (especially Emilia Rubio) for providing the abstract similarity assessment that made the first filtering step possible.  Finally, we thank Anthony Opipari and Marc Pujol-Gonzalez for critical behind-the-scenes assistance with the logistics of communicating decisions about dual submissions.

\appendix
\section{LLM Overlap Checker Prompt Construction}
\label{app:LLM}

The production overlap checker used GPT-5 through the OpenAI Responses API with structured parsing into two fields: \texttt{severity} and \texttt{explanation}.  The model was called with high reasoning effort, and the batch script retried transient failures before recording a failed comparison.  The prompt did not include the OpenReview title/abstract similarity score or author metadata.
\newpage
\begin{verbatim}
instructions = (
    "You will be given two papers that were submitted to the same conference. Evaluate 
    both of them, and let me know if they are malicious attempts by the authors to thwart
    dual submission policies or otherwise unethically game the system to get at least one 
    paper accepted."
    "Return a JSON object with fields 'severity' (integer 0-5) and 'explanation' (string). 
    The severity scale meanings are:\n"
    "0: No overlap at all;\n"
    "1: Minimal conceptual overlap;\n"
    "2: Some shared ideas but no significant overlap in methodology or results;\n" 
    "3: Moderate overlap with common notation, terminology, or acronyms;\n"
    "4: Extensive overlap with strong indicators of coordinated effort that should have 
    been reported;\n"
    "5: Clear plagiarism, with verbatim reuse of text, methodology, or results."
)
messages = [
    {"role": "developer", "content": [RM.input_text_message(instructions)]},
    {
        "role": "user",
        "content": [
            RM.input_text_message("<submission_a PDF>\n"),
            RM.input_file_id(paper1_id),
            RM.input_text_message("</submission_a PDF>\n\n"),
            RM.input_text_message("<submission_b PDF>\n"),
            RM.input_file_id(paper2_id),
            RM.input_text_message("</submission_b PDF>\n\n"),
        ],
    },
]
\end{verbatim}

\bibliographystyle{plainnat}  
\bibliography{references}  

\end{document}